\documentstyle[preprint,aps]{revtex}
\begin{document}
\draft
\title{Chaos around a H\'enon-Heiles-inspired exact perturbation
 of a black hole}
\author{Werner M. Vieira\thanks{e-mail: vieira@ime.unicamp.br} and
 Patricio S. Letelier\thanks{e-mail: letelier@ime.unicamp.br}}
\address{Departamento de Matem\'atica Aplicada\\
Instituto de Matem\'atica, Estat\'{\i}stica e Ci\^encias da
 Computa\c{c}\~ao\\
Universidade Estadual de Campinas, CP 6065\\13083-970
 Campinas, SP, Brazil
\\{\rm January 10, 1996 }}
\maketitle
\begin{abstract}
A solution of the Einstein's equations that represents the
 superposition of
 a Schwarszchild black hole with both quadrupolar and octopolar
 terms
 describing a halo is exhibited.  We show that this solution, 
in the Newtonian limit, is an
 analog to the well known H\'enon-Heiles potential.
 The integrability of orbits of test particles moving around a
 black hole representing the galactic center is studied and
 bounded zones of chaotic behavior are found.\\

\end{abstract}


There are two main lines of research of chaotic behavior in General
 Relativity: one
 deals with chaos associated to the existence of generic
 singularities in
 homogeneous cosmological models, as in Bianchi IX 
model~\cite{Belinski}. In 
this case, Einstein's equations themselves -- although suitably 
reduced to a 
simple toy model -- form the dynamical system being considered. 
Invariance
 of the relativistic theory against arbitrary coordinate changes, 
particularly 
time changes, challenges the proper concept of chaos in 
this approach~\cite{FranciscoMatsas}.
The other line assumes a given geometry and looks for chaotic 
behavior of
 geodesic motion in this background. In this case, the geometry
 can be taken 
as either an approximate 
or an exact solution of Einstein's equations. Examples of chaotic 
geodesic 
motion in exact geometries are considered 
in~\cite{Contopoulos,JAP,KarasVokrouh}. 
On the other hand, a linearized treatment of the geometry is made
in~\cite{BombelliCalzetta} and~\cite{Moeckel}, to model the 
interaction of a 
black hole with a weak gravitational wave and a weak gravitational 
dipole 
moment, respectively. 

Firstly in this letter we consider -- with some astronomical 
motivation --
 an  exact solution of Einstein's equations that represents a 
Schwarszchild black hole together with quadrupolar and octopolar 
contributions of arbitrary strength. Since the multipolar 
contribution can be switched off, this solution is considered to 
represent a perturbation of a black hole. Secondly, we 
address the question whether
  this exact perturbation breaks the integrability of test
 motions around the black
 hole. Some previous results~\cite{BombelliCalzetta,Moeckel}
 deal with
 linearized perturbations of the Schwarszchild background which, 
although 
interesting, make the perturbation ``strong enough'' to push
 it out of
 the full theory. This is an important  limitation since
 we know that 
chaos is rather the rule than the exception in the general
 context of 
dynamical systems and that only a small change in the dynamics 
could drastically alter the motion portrait.

These facts motivate us to look for a solution of
 vacuum Einstein's 
equations whose additional controlable parameters are linked
 to multipolar structures, 
originated in the model from exterior matter distributions.
 To this end, 
our search has been inspired by the well
 known H\'enon-Heiles potential (HHP) -- classical in
 astronomy and also
 in dynamical systems--  one of the most studied paradigms of 
chaotic
 behavior~\cite{HH,Fordy,Berry}. It was  originally intended to 
model a
  Newtonian axially symmetric galactic potential. In fact,
 we have found
 a solution 
that is a general relativistic analog  of HHP and
 moreover seems to be
 a less idealized model for an axially symmetric
 galaxy with a massive 
core plus an exterior halo of quadrupoles and octopoles.

Finally, we shall see that the quadrupolar terms
 only are not sufficient to
 cause chaotic geodesic motion. The chaotic behavior
 occurs only due to the
 presence of the octopolar terms, which break the
 reflection symmetry
 of the potential about the middle plane of the galaxy,
 in the same manner
that the cubic terms in the HHP give rise to the
 nonintegrability of the motion.

We start with the Weyl metric describing static axially
 symmetric space-times
\begin{equation}
ds^2=e^{2\nu}c^2dt^2-e^{-2\nu}[e^{2\gamma}(dz^2+dr^2)
+r^2d\phi^2].\label{hh1}
\end{equation}
$\nu$ and $\gamma$ are functions of $r$ and $z$ only.
 The Einstein's equations 
reduce in this case  to the usual Laplace equation
 $ \nu_{,rr}+\nu_{,r}/r+\nu_{,zz}=0$ and the quadrature 
$d\gamma=r[(\nu_{,r})^2-(\nu_{,z})^2]dr+2r\nu_{,r}\nu_{,z}dz$. 
Despite the
 simplicity of the Einstein's equations in the
 coordinates $r$ and $z$, they 
are rather deceiving~\cite{PBH}; better coordinates are the
 prolate 
spheroidal $u$ and $v$ 
defined as
\begin{eqnarray}
 u&=&\frac{1}{2L}[\sqrt{r^2+(z+L)^2}+\sqrt{r^2+(z-L)^2}]
{\mbox{,\hspace{0.4cm}$u \geq 1,$}}\nonumber\\
v&=&\frac{1}{2L}[\sqrt{r^2+(z+L)^2}-\sqrt{r^2+(z-L)^2}]{\mbox{,
\hspace{0.4cm}$-1\leq v\leq 1$}},\label{hh4}
 \end{eqnarray}
where $L$ is an arbitrary constant. In these coordinates
 Laplace's
 equation can be solved in terms of standard Legendre
 polynomials. In particular, we are interested in the solution
\begin{equation}
\nu(u,v)=a_0Q_0(u)+b_2P_2(u)P_2(v)+b_3P_3(u)P_3(v).\label{hh8}
\end{equation}
The first term is the monopole one, which is related
 to the Schwarszchild
 solution with $a_0=-1$ and $L=GM/c^2=R_0$.
 The remaining terms represent
 the multipolar structure of a halo. We impose
 the additional condition
 of elementary flatness over $r=0$ for $|z| > L$,
 i.e.\ , $\gamma(u,v)=0$ in
 this region of the $z$ axis, thus eliminating conical 
singularities there.

The integration of $\gamma$ is guaranteed by the fact that
 $\nu$ is a solution 
of Laplace's equation.  We find
\begin{eqnarray}
2\nu&=&(1+\epsilon)\log(\frac{u-1}{u+1})+\nu_Q(u,v)
+\nu_O(u,v),\nonumber\\
2\gamma&=&(1+\epsilon)^2\log(\frac{u^2-1}{u^2-v^2})
+\gamma_Q(u,v)+
\gamma_O(u,v)+\gamma_{QO}(u,v),\label{hh9}
\end{eqnarray}
where $\epsilon$ is an arbitrary constant and
\begin{equation}
\begin{array}{l}
\nu_Q=({\cal{Q}}/3)(3u^2-1)(3v^2-1),\\
\nu_O=({\cal{O}}/5)uv(5u^2-3)(5v^2-3),\\
\gamma_Q=-4{\cal{Q}}(1+\epsilon)u(1-v^2)+({\cal{Q}}^2/2)
[9u^4v^4-10u^4v^2-10u^2v^4\\
{\mbox{\hspace{1.0cm}}}+12u^2v^2+u^4+v^4-2u^2-2v^2+1],\\
\gamma_O=-2{\cal{O}}(1+\epsilon)[3u^2v-3u^2v^3+v^3
-\frac{9}{5}v+\frac{4}{5}]
+2{\cal{O}}^2[ \frac{75}{8}u^6v^6 -\frac{117}{8}u^6v^4\\
{\mbox{\hspace{1.0cm}}}-\frac{117}{8}u^4v^6 +\frac{45}{8}u^6v^2 
+\frac{45}{8}u^2v^6 +\frac{189}{8}u^4v^4 -\frac{387}{40}u^4v^2
 -\frac{387}{40}u^2v^4 +\frac{891}{200}u^2v^2\\
{\mbox{\hspace{1.0cm}}}-\frac{3}{8}u^6 -\frac{3}{8}v^6
 +\frac{27}{40}u^4 +\frac{27}{40}v^4 -\frac{81}{200}u^2 
-\frac{81}{200}v^2
 +\frac{21}{200}],\\
\gamma_{QO}=2{\cal{Q}}{\cal{O}}[9u^5v^5 -12u^5v^3 +3u^5v
-12u^3v^5 +\frac{84}{5}u^3v^3 -\frac{24}{5}u^3v +3uv^5 
-\frac{24}{5}uv^3
 +\frac{9}{5}uv].\label{hh11}
\end{array}
\end{equation}
${\cal{Q}}$ and ${\cal{O}}$ are respectively the quadrupole
 and the octopole strengths. 
The terms proportional to ${\cal{Q}}$ and ${\cal{O}}$ in
 $\gamma_Q$ and $\gamma_O$, 
respectively, represent  nonlinear interactions between
 the black hole
 and the multipoles. 
For $\epsilon =0$ and $L=R_0$ we have the Schwarszchild solution
 perturbed by quadrupoles and octopoles and,
 if $\epsilon = -1$, we are
 left only with the multipolar structure. Note that only the odd
 terms in $v$ break the reflection symmetry
 about the plane $z=0$ and
 all them are linked only to the octopole
 strength. The particular case 
 $\epsilon ={\cal{ O}} =0$ and $L=R_0$ is considered in~\cite{PBH}.

To see that the multipolar structure in the solution
 (\ref{hh9}) and (\ref{hh11})
is a general relativistic analog of HHP, we write the
 solution in the usual Schwarszchild coordinates 
$(t,\rho,\theta,\phi)$
\begin{equation}
z=(\rho-L)\cos\theta,{\mbox{\hspace{0.4cm}}}r=\sqrt{\rho(\rho-2L)}
\sin\theta.\label{hh12}
\end{equation}
For $\epsilon=0$ and $L=R_0$ in (\ref{hh4}) plus
 (\ref{hh12}), we get

\begin{eqnarray}
ds^2=(1-\frac{2R_0}{\rho})e^{(\nu_Q+\nu_O)}c^2dt^2-e^{(\gamma_Q+
\gamma_O+\gamma_{QO}-\nu_Q-\nu_O)}[(1-\frac{2R_0}{\rho})^{-1}
d\rho^2+
\rho^2d\theta^2]\nonumber\\
{\mbox{\hspace{8.5cm}}}+e^{-(\nu_Q+\nu_O)}\rho^2\sin^2\theta 
d\phi^2.\label{hh13}
\end{eqnarray}

In the Newtonian approximation only $g_{tt}$ in (\ref{hh13}) is 
important. We expand it to the first order in the strengths
 ${\cal{Q}}$ and ${\cal{O}}$ and identify a Newtonian
 potential $\Phi_N$ 
through
\begin{equation}
g_{tt}=1+\frac{2}{c^2}\Phi_N{\mbox{\hspace{0.4cm}.}}\label{hh14}
\end{equation}
Remember that (\ref{hh12}) are valid only for $\rho\geq 2R_0$.
 If we 
suppose that there exists some region between the horizon
 event $\rho=2R_0$ of
 the galactic core and the external multipolar structure
 simulating the halo, in
 which the Schwarszchild
 coordinates are approximately the Euclidean spherical
 ones, then there we 
have $z\approx \rho\cos\theta$ and $\rho^2\approx z^2+R^2$
 with $R$ being
 the radial cylindrical coordinate. Expanding (\ref{hh14})
 and using 
this approximation, we arrive at
\begin{equation}
\Phi_N=-\frac{GM}{\rho}+\frac{{\cal{Q}}c^2}{2R_0^2}[2z^2-R^2
+f(\rho,z)]\\
+\frac{{\cal{O}}c^2}{2R_0^3}[2z^3-3zR^2+g(\rho,z)]
{\mbox{\hspace{0.4cm},}}\label{hh15}
\end{equation}
where
\begin{eqnarray}
f(\rho,z)&=&-\frac{R_0^2}{3}(3\frac{z^2}{\rho^2}-1)(12
\frac{\rho}{R_0}-14+4\frac{R_0}{\rho}),\nonumber\\
g(\rho,z)&=& -\frac{R_0^2}{5}z(5\frac{z^2}{\rho^2}-3)
(25\frac{\rho}{R_0}-42+26\frac{R_0}{\rho}-4\frac{R_0^2}{\rho^2}).
\label{hh16}
\end{eqnarray}

For comparison we quote HHP from~\cite{HH}
\begin{equation}
\Phi_{HH}=\frac{1}{2}[z^2+R^2+2zR^2-\frac{2}{3}z^3].\label{hh17}
\end{equation}

The quadrupolar and octopolar terms explicitly
 shown in (\ref{hh15}) are 
qualitatively the same as those appearing in HHP,
 with one difference
 that will be of importance in the latter discussion:
 in HHP the quadrupolar
 terms constitute an (isotropic) bidimensional oscilator,
 while in our case
 we have a saddle point. Once HHP is considered to be
 by far more interesting by its
 dynamical richness allied to simplicity than by
 actually modeling any
 concrete astronomical situation, the similarity
 above justifies the 
parallelism. Moreover, the quadratic and cubic
 terms in (\ref{hh15}) are
 indubitably in the {\it{extended}} H\'enon-Heiles
 family of potentials~\cite{Fordy}.

In Fig.~\ref{fig1} we show Poincar\'e sections
 across the plane $z=0$
 in Weyl coordinates for some time-like geodesic
 orbits for the metric
 (\ref{hh9}) and (\ref{hh11}). In Fig.\ 1(a) we
 show a section for the
 unperturbed Schwarszchild background ($\epsilon=0$, 
${\cal{Q}}={\cal{O}}=0$), where
 the complete integrability is seen.
 In Fig.\ 1(b) we switch on the octopolar
 terms over Fig.\ 1(a) ($\epsilon=0$, ${\cal{Q}}=0$
 and ${\cal{O}}\neq 0$). This
 breaks the reflection symmetry about the middle
 plane and hence also the
 integrability of the motion. We also have checked that 
this chaotic manifestation
 is not explained by the signal structure of the eigenvalues
 of the curvature tensor as proposed in~\cite{JAP}.
 We see that the octopole terms
 are a powerful source of chaos since the strength we have
 used is only ${\cal{O}}=1.5\times 10^{-7}$ in nondimensional 
units. In Fig.\ 1(c) we switch on the quadrupolar terms
 over Fig.\ 1(a) ($\epsilon=0$, ${\cal{Q}}\neq 0$ and 
${\cal{O}}=0$). We observe
 that integrability is preserved, yet in a more
 involved toroidal structure
 in phase space. Finally, Fig.\ 1(d) exhibits
 chaotic behavior due to the
 full perturbation of the Schwarszchild background
 ($\epsilon=0$, ${\cal{Q}}\neq 0$ and ${\cal{O}}\neq 0$).
 We note
 that although the quadrupolar terms do not cause
 chaos, their presence
 only with strength ${\cal{Q}}=2.2\times 10^{-6}$ does the system
 more sensible to the octopolar terms.

By switching the black hole off ($\epsilon=-1$)
 -- with or without the octopoles -- we 
were not able to find  regions of bounded motion
 that include points outside
 the middle plane. This lack of nonplanar bounded
 regions when solely the multipolar
 halo is taken into account is probably related to the 
{\it{repulsive}} quadractic $z$-term present in the
 expansion (\ref{hh15}).

The authors thank  CNPq and FAPESP for financial support.

\begin{figure}

\caption{Poincar\'e sections (in Weyl coordinates)
 in the plane $z=0$ for
 timelike geodesic orbits in different geometries.
 All test particles
 (with mass $\mu$) have energy $E=0.975\mu c^2$ and
 angular momentum 
$L=3.8R_0\mu c$. In all figures $\epsilon=0$ and 
the nondimensional 
variables are defined by $X\equiv r/R_0$ 
and $dT\equiv ds/R_0$: (a)  
unperturbed Schwarszchild metric (${\cal{Q}}={\cal{O}}=0$). 
(b) Schwarszchild metric perturbed by octopolar 
terms (${{\cal{Q}}=0,\cal{O}}=1.5\times  10^{-7}$).
 (c) Schwarszchild
 metric perturbed by quadrupolar terms
 (${{\cal{Q}}=2.2\times  10^{-6},\cal{O}}=0$). 
(d) Schwarszchild metric plus the full perturbation 
(${{\cal{Q}}=2.2\times 
 10^{-6},\cal{O}}=1.5\times  10^{-7}$).}\label{fig1}

\end{figure}

\end{document}